# Predicting the longevity of resources shared in scientific publications

**Daniel E. Acuna**[1]    **Jian Jian**[1]    **Tong Zeng**[2]    **Lizhen Liang**[1]    **Han Zhuang**[1]

[1]School of Information Studies
Syracuse University
[2]School of Information Management
Nanjing University

**Abstract**

Research has shown that most resources shared in articles (e.g., URLs to code or data) are not kept up to date and mostly disappear from the web after some years (Zeng et al., 2019). Little is known about the factors that differentiate and predict the longevity of these resources. This article explores a range of explanatory features related to the publication venue, authors, references, and where the resource is shared. We analyze an extensive repository of publications and, through web archival services, reconstruct how they looked at different time points. We discover that the most important factors are related to *where* and *how* the resource is shared, and surprisingly little is explained by the author's reputation or prestige of the journal. By examining the places where long-lasting resources are shared, we suggest that it is critical to disseminate and create standards with modern technologies. Finally, we discuss implications for reproducibility and recognizing scientific datasets as first-class citizens.

# Introduction

Scientific research produces not only publications but also datasets and computer code. These artifacts have become increasingly crucial for reproducibility and replication. Worryingly, recent research has shown that resources shared in scientific articles tend to disappear over time. For example, a recent study estimated that more than half of the URLs inside articles disappear after eight years (Zeng et al., 2019). However, it is unclear which factors determine the disappearance of resources. Not knowing these factors might not allow us to correct the situation. Here, we propose analyzing authors, the journals, and the technology behind sharing information to understand this association. Our results will attempt to illuminate this underdeveloped area of scientific resources beyond publications.

Sharing resources has always been an essential part of the scientific process. For example, it can help increase the reproducibility of results of other studies (Vasilevsky et al., 2013), allowing fellow scientists to build upon previous work more efficiently. In addition, replicability is a more vital form of scientific correctness check, requiring an entirely new group to redo and confirm experimental results (National Academies of Sciences & Medicine, 2019). Thus, in all these settings, sharing resources is essential.

Historically, it has been expensive to share resources in scientific work. Also, once resources are shared, they might need to be kept available for an extended period. Several studies have revealed that resources are lost surprisingly early and frequently. A recent study on a dataset of resources shared in open access publications estimated that more than half of them are lost within eight years after publication (Zeng et al., 2019). Some researchers have speculated about the factors that make resources be lost (Zhou et al., 2015). One of the issues with many studies is that their scale is small, and they cannot analyze *when* resources are lost—they can only get the view of the resources at the time of the particular study. Sharing resources in science has become more manageable, but little is understood about why their upkeep fails across time.

In this study, we enrich the data from (Zeng et al., 2019), and reconstruct how the resources and publications would have looked like at the time of publication. In this manner, we know precisely *when* a resource disappeared from the web. Also, we compute features that we hypothesize are predictive of when a resource disappears, including the type of technology used for sharing, the prestige of the authors, and publication venues. We built two statistical models for predicting resources longevity: one highly interpretable based on a censored model and one highly predictive but less interpretable based on Random Forests. Our analysis reveals that the technology where the resources are stored is essential for predicting resource longevity. Surprisingly, we find that the prestige of the author and institution are almost irrelevant. We discuss the implications of these results related to how we can plan to keep resources available in the future.

## Literature review

There is a wide range of research shared in articles. The most obvious ones are shared within the article itself: figures, tables, statistical analysis, and claims. We do not think of these resources as "decaying" as they are contained within the article. If we have mechanisms for storing the article, then we can recover these resources perfectly. However, we cannot share all resources in an article. Articles sometimes rely on big datasets that are shared on personal websites of the resources or external services. The datasets, in turn, could be of several kinds, including genomic sequences, machine learning datasets, images, text, video, audio, among others. Other artifacts that are especially common these days are software code, preprocessing scripts, and binary code. Some of these resources are sometimes shared with their entire history (e.g., Github) or with highly reproducible container environments (e.g., Docker image). Regardless, science is more and more dependent on datasets and code that are widely shared.

The Internet is a significant facet of modern science and a crucial research tool to store and share resources in place of libraries and physical storage (Lawrence & Giles, 2000). Thus, network-accessible information is increasingly being cited by journal articles (Duda & Camp, 2008). While scholarly articles on the Internet are referenced, assets used or created during research like software, ontologies, and datasets are also referenced (Klein et al., 2014). Yet resources from the Internet can be unreliable due to the dynamic nature of the web (Duda & Camp, 2008). Multiple researchers have found a consistent lack of persistence in these resources (Klein & Nelson, 2010; Koehler, 1999, 2004). Thus, a significant issue is maintaining the value of scientific knowledge using storage systems and structures that are fragile.

Predictably, pany researchers have found that Internet-stored resources cited by scientific articles suffer from "reference rot" or "link rot" (Klein et al., 2014; Van de Sompel et al., 2014). More specifically, the resource identified by a URI ceases to exist and becomes unavailable. A related phenomenon is "content drift," where resources shared in URIs change over time and cease to be relevant for what was initially referenced. The field of biomedical science has been a pioneer in attempting to avoid these issues but still has some problems in software (Mangul et al., 2018), protocols (Open Science Collaboration, 2015), and datasets (Bonàs-Guarch et al., 2018). Due to the importance and cost of biomedical science, the fact that these resources disappear should be of great concern. We might expect that other fields considered less crucial suffer similar or worse problems. Therefore, we need mechanisms to keep track of this disappearance, drifting, and irrelevance.

Government and funding agencies have been pushing to establish higher standards for data sharing. Around 2003, the National Institutes of Health (NIH) published a policy requiring applications for grants greater than $500,000 to include data-sharing plans (National Institutes of Health, 2003). The National Science Foundation has developed similar policies encouraging data sharing (National Science Foundation, 2011). Countless other institutions recognize the importance of such practice (e.g., see (Milham et al., 2018; Van Horn & Gazzaniga, 2013)) and its impact on science (e.g., see (Bonàs-Guarch et al., 2018)). Sharing of resources is essential, and we need to better understand the process of its decay.

### *Why predicting resource longevity might be helpful.*

Ideally, we would like to predict how resources decay to optimally allocate when to perform maintenance or simply decide that a resource is no longer relevant. Many systems work based on this principle. For example, libraries with long-term storage systems need to understand how storage decays and therefore need to plan accordingly to update storage systems, ventilation, and other factors (El-Bakry & Mohammed, 2009). Furthermore, in the digital world, long-term storage requires engineers to plan backups, upgrades, and removal policies to ensure a good tradeoff between storing relevant information accurately and removing information that is not used.

In science, we do not think about these issues directly. However, we do share our findings as soon as possible, indirectly implying that resources are most useful sooner rather than later. For example, when we cite previous research, it is standard practice to cite relevant publications recently published (Penders, 2018). The understanding is that research published too long in the past may not translate well to today's research questions (Plavén-Sigray et al., 2017). When publications share software, this is even more apparent. In machine learning, for example, it is common practice that the code and data are shared alongside the software and data storage versions so that other scientists in the future can reproduce the results (Kop, 2020). While research has shown that machine learning might lack reproducibility (Haibe-Kains et al., 2020), machine learning is a field where sharing artifacts beyond publications is standard practice at least. Therefore, scientists, either implicitly or explicitly, are always thinking about the relevance of resources.

## Materials and methods

*Data*

*Dataset of URLs in Open Access publications*

In this research, we enrich data from (Zeng et al., 2019). That study collected the URL links within PubMed Open Access Subset (PMOAS) publications and checked whether they were accessible. The study extracted several features from the authors, publications, and the URLs contained therein, including domain country, size of the resource, number of affiliations, *h*-index of the journal, age of the URL, number of references, length of the abstract, and the frequency of use of URL in other publications. The study found that the frequency of reuses and the "int" (international), "org" (mostly non-profit organizations), and "gov" (governmental) domains are most predictive of a link being alive. Conversely, it found that the length of the URL's path (after the domain), and the Indian, European Union, and Chinese domains were most predictive of resources not being accessible. However, this study did not investigate how long a project would last and did not use features related to citations, the technology used, and the prestige of the authors.

*Dataset enrichment*

Here we aim to predict the longevity of resources based on the features present when authors first share the resources. The features we look at are not just related to the resource itself, but the technology used and the publication, authors, and venue. We also use the Microsoft Academic Graph (MAG) to include relationships between publications, authors, institutional affiliations, and citations. With the dataset, we are able to generate features inferred from those relationships that are helpful to the prediction task.

To reconstruct how the resources (i.e., URLs) looked at the time of publication, we have to use web archiving services. In particular, we use the Wayback Machine API[1], which allows us to access how a URL resource was during an arbitrary point in the past. We filter out several kinds of URLs. First, we ruled out those URLs which represent FTP resources, so we can focus on analyzing the quality of the website. We also filter out URLs that are not available in the original destination or not archived in the Wayback Machine. We logged the last available date for all of the rest URLs. If we scraped the URL from its original location, then the current timestamp would be recorded as its last available time. Otherwise, we filled it with the archive's timestamp representing the last time it appeared in the Wayback Machine. Finally, we extracted information from the HTML's source code and dependencies. In particular, we extract features of the technical structure, such as the presence of the iframe, the type of the suffix and charset, and the number of internal and external JavaScript dependencies.

We noticed that a large proportion of our URLs are reachable today; about 90% of the URLs were alive when we fetched them in April of 2020. Their lifespan could be more than the age calculated by subtracting it from the year it first becomes available.

---

[1] https://archive.org/web/

*Methods*

*Feature engineering*

Considering that we are trying to predict the longevity of a resource shared in a scientific article, in this section we explain the features we extract from the URL itself, the website's HTML and resources, and authors of the publication, the institution, the journal, and the citation structure of the article itself. This feature engineering resulted in 42 independent variables and 1 dependent variable (Table. 1). We explain in more detail the rationale for these features in the Supplementary Material Table S1.

**Table 1. Features and their rationale.** Some of the features are borrowed from (Zeng and Acuna, 2019). More detailed explanation in Table S1.

| Feature group | Feature set |
| --- | --- |
| **URL information** | **1)** protocol type; **2)** depth of the URI path; **3)** www or non-www URI; **4)** level of subdomain; **5)** number of query parameters in the URI; domain suffixes: **6)** .ORG, **7)** .INT, **8)** .JP, **9)** .GOV, **10)** .IN, **11)** .CN, **12)** .EU, and **13)** .KR; **14)** port number used |
| **HTML contents and technology** | **15)** size of the HTML source code; **16)** length of HTML title tag; **17)** number of internal (same server) Javascript files; **18)** number of external Javascript files; **19)** charset specified in the HTML source code; **20)** HTML or HTML5; **21)** iframe used; **22)** number of hyperlinks used; |
| **Article references** | **23)** # references; **24)** # authors referenced; **25)** # affiliations referenced; **26)** # journals referenced; **27)** # author self-referenced; **28)** # affiliation self-referenced; **29)** # journal self-referenced; **30)** Average year of publications referenced; **31)** Minimum, **32)** maximum, and **33)** median year of publications referenced; |
| **Authors quantity** | **34)** # authors; **35)** number of affiliations in the paper; **36)** # authors' citation; **37)** total number of authors' citation(may be duplicated); **38)** total number of affiliations' citation (may be duplicated) |
| **Authors' prestige/seniority/impact** | **39)** Average h-index of author(s); **40)** H-index of the first author; **41)** H-index of the last author; **42)** Average h-index of middle author(s) |
| **Dependent variable** | |
| Last known year of the website | **43)** Last available year of resource |

*Models*

We will use two kinds of models for our modeling and prediction. One model which is interpretable and has amenable statistical properties is based on a censored regression model. The other model is based on Random Forest.

*Censored model of resource decay*

Most of the URLs we analyze are still available, and therefore, we do not know when they will become unavailable. However, we want to use this information to inform the model that predicts longevity. To do this task, we will use techniques from censored regression models. These types of regression models split the prediction into two parts: one predicts whether the resource is alive or not, and the other predicts how long it will take to perish if the resource is predicted to be alive.

In particular, we will use a Tobit model to perform a censored regression. We now explain in detail how such a model fit works using a Bayesian maximum aposteriori framework. To avoid excessive overfitting, we impose a prior distribution over the parameters and maximize the posterior distribution of the parameters rather than simply the likelihood function. Our Tobit model constructs a data likelihood $L(\beta, \sigma)$ based on the parameters $\beta$ and $\sigma$ and weighs that by the prior distribution $P(\beta)$ over the parameters. Concretely, the likelihood function is given by

$$L(\beta, \sigma) = \prod \left[ \Phi\left(\frac{a - X^T\beta}{\sigma}\right) \right]^{I^a}$$

$$\times \prod \left[ \frac{\varphi\left(\frac{y - X^T\beta}{s}\right)}{\sigma} \right]^{(1 - I^a - I^b)}$$

$$\times \prod \left[ \Phi\left(\frac{X^T\beta - b}{\sigma}\right) \right]^{I^b}$$

and we assume that the prior distribution follows a mixture of Laplacian distribution and Gaussian distribution, which is similar to assuming L2 and L1 regularization parameters as in Elastic Net regularization:

$$P(\beta) = \left( f_{Gaussian}(\beta)^{(1-\alpha)} \cdot f_{Laplacian}(\beta)^{\alpha} \right)^{\lambda}$$

Then the posterior distribution is given by

$$posterior = \frac{likelihood \cdot prior}{evidence}$$

$$P(\beta \mid Y) = \frac{P(Y \mid \beta) \cdot P(\beta)}{P(Y)}$$

$$= \frac{L(\beta, \sigma) \cdot P(\beta)}{P(Y)}$$

$$P(\beta \mid Y) = \frac{L(\beta, \sigma) \cdot \left( f_{Gaussian}(\beta)^{(1-\alpha)} \cdot f_{Laplacian}(\beta)^{\alpha} \right)^{\lambda}}{P(Y)}$$

The log-posterior distribution will then take the more amenable optimization form

$$\log(P(\beta \mid Y)) = \log\left(\frac{L(\beta,\sigma) \cdot f_{Gaussian}(\beta)^{(1-\alpha)} \cdot f_{Laplacian}(\beta)^{\alpha}}{P(Y)}\right)$$

$$\begin{aligned}\log(P(\beta \mid Y)) &= \log(L(\beta,\sigma)) \\ &\quad + \log\left(f_{Gaussian}(\beta)^{(1-\alpha)\cdot\lambda}\right) \\ &\quad + \log\left(f_{Laplacian}(\beta)^{\alpha\cdot\lambda}\right) \\ &\quad - \log(P(y)) \\ &= \log(L(\beta,\sigma)) \\ &\quad + \lambda \cdot (1-\alpha) \cdot \log(f_{Gaussian}(\beta)) \\ &\quad + \lambda \cdot \alpha \cdot \log(f_{Laplacian}(\beta)) \\ &\quad - \log(P(Y))\end{aligned}$$

As we estimate the probability in terms of the β, we can eliminate the constant term evidence and apply MLE to it.

$$\hat{\beta} = \arg\max P(\beta \mid Y)$$

$$\begin{aligned} &= \arg\max \log(P(\beta \mid Y)) \\ &= \arg\max \log(L(\beta \mid Y)) \\ &\quad + \lambda \cdot (1-\alpha) \cdot \log(f_{Gaussian}(\beta)) \\ &\quad + \lambda \cdot \alpha \cdot \log(f_{Laplacian}(\beta)) \\ &\quad - \log(P(Y)) \\ &= \arg\max \log(L(\beta,\sigma)) \\ &\quad + \lambda \cdot (1-\alpha) \cdot \log(f_{Gaussian}(\beta)) \\ &\quad + \lambda \cdot \alpha \cdot \log(f_{Laplacian}(\beta)) \end{aligned}$$

We will use an L-BFGS-B optimization which requires to provide the gradient, which will simply become

$$\frac{\Delta \log P(\beta \mid \sigma)}{\Delta \beta} = \frac{\Delta \log(L(\beta,\sigma)) + \lambda \cdot (1-\alpha) \cdot \log(f_{Gaussian}(\beta)) + \lambda \cdot \alpha \cdot \log(f_{Laplacian}(\beta))}{\Delta \beta}$$

$$\frac{\Delta \log P(\beta \mid \sigma)}{\Delta \beta}$$

$$= \frac{\Delta \log(L(\beta, \sigma))}{\Delta \beta} + \frac{\lambda \cdot (1-\alpha) \cdot \log(f_{Gaussian}(\beta))}{\Delta \beta}$$

$$+ \frac{\lambda \cdot \alpha \cdot \log(f_{Laplacian}(\beta))}{\Delta \beta}$$

For a Gaussian distribution:

$$\frac{\log(f_{Gaussian}(x))}{\Delta x} = \frac{\log\left(\frac{1}{\sigma \cdot \sqrt{2 \cdot \pi}} e^{-\frac{1}{2} \cdot \left(\frac{x-\mu}{\sigma}\right)^2}\right)}{\Delta x}$$

$$= \frac{\log\left(\frac{1}{\sigma \cdot \sqrt{2 \cdot \pi}}\right) + \log\left(e^{-\frac{1}{2} \cdot \left(\frac{x-\mu}{\sigma}\right)^2}\right)}{\Delta x}$$

$$= \frac{\log\left(\frac{1}{\sigma \cdot \sqrt{2 \cdot \pi}}\right)}{\Delta x} + \frac{\log\left(e^{-\frac{1}{2} \cdot \left(\frac{x-\mu}{\sigma}\right)^2}\right)}{\Delta x}$$

$$= \frac{\log\left(e^{-\frac{1}{2} \cdot \left(\frac{x-\mu}{\sigma}\right)^2}\right)}{\Delta x}$$

$$= \frac{-\frac{1}{2} \cdot \left(\frac{x-\mu}{\sigma}\right)^2}{\Delta x}$$

$$= -\frac{x}{\sigma}$$

We assume that the standard deviation in the prior is a unit value, further simplifying it into

$$\frac{\log(f_{Gaussian}(x))}{\Delta x} = -x$$

Similarly, for the Laplacian distribution

$$\frac{\log(f_{Laplacian}(x))}{\Delta x} = \frac{\log\left(\frac{1}{2b} e^{-\frac{|x-\mu|}{b}}\right)}{\Delta x}$$

$$= \frac{\log\left(\frac{1}{2b}\right) + \log(e^{-\frac{|x-\mu|}{b}})}{\Delta x}$$

$$= \frac{\log\left(\frac{1}{2b}\right)}{\Delta x} + \frac{\log\left(e^{-\frac{|x-\mu|}{b}}\right)}{\Delta x}$$

$$= \frac{\log\left(e^{-\frac{|x-\mu|}{b}}\right)}{\Delta x}$$

$$= \frac{-\frac{|x-\mu|}{b}}{\Delta x}$$

By applying the subgradient, we get the expression below

$$\frac{\log\left(f_{Laplacian}(x)\right)}{\Delta x} = -1 \;\; if \; x > 0$$

$$\frac{\log\left(f_{Laplacian}(x)\right)}{\Delta x} = 0 \;\;\;\; if \; x = 0$$

$$\frac{\log\left(f_{Laplacian}(x)\right)}{\Delta x} = 1 \;\;\;\; if \; x < 0$$

Let the functions $L1(\beta)$ and $L2(\beta)$ be

$$L1(\beta) = \{-1 \; if \; x_i > 0; \; 0 \; if \; x_i = 0; \; 1 \; if \; x_i < 0\}$$
$$L2(\beta) = X$$

we can express the gradient of log a posteriori as

$$\frac{\Delta \log(L(\beta,\sigma))}{\Delta \beta}$$

$$= \sum_{i=1}^{N} \left[ -I^a \frac{\varphi\left(\frac{a - X^T\beta}{\sigma}\right)}{\Phi\left(\frac{a - X^T\beta}{\sigma}\right)} \frac{x}{\sigma} + I^b \frac{\varphi\left(\frac{X^T\beta - b}{\sigma}\right)}{\Phi\left(\frac{X^T\beta - b}{\sigma}\right)} \frac{X}{\sigma} \right.$$

$$+ (1 - I^a - I^b)\frac{Y - X^T\beta}{\sigma}\frac{X}{\sigma}$$

$$\left. + \lambda \cdot \left(\alpha \cdot L1(\beta) + (1 - \alpha \cdot)L2(\beta)\right)\right]$$

$$\frac{\Delta \log(L(\beta,\sigma))}{\Delta \log(\sigma)}$$

$$= \sum_{i=1}^{N} \left[ -I^a \frac{\varphi\left(\frac{a - X^T\beta}{\sigma}\right)}{\Phi\left(\frac{a - X^T\beta}{\sigma}\right)} \frac{a - X^T\beta}{\sigma} + I^b \frac{\varphi\left(\frac{X^T\beta - b}{\sigma}\right)}{\Phi\left(\frac{X^T\beta - b}{\sigma}\right)} \frac{X^T\beta - b}{\sigma} \right.$$

$$\left. + (1 - I^a - I^b)\left(\left(\frac{Y - X^T\beta}{\sigma}\right)^2 - 1\right) \right]$$

We use the above gradients together with the Python package `scipy.optimize` to find the optimal parameters. Because we needed to estimate the uncertainty about these parameters, we use bootstrapping estimation of the posterior distribution (Murphy, 2012).

*Random Forest model*

The Tobit model above has the important property that it is easy to understand through the $\hat{\beta}$ features: they are a simple regression that will linearly influence how long a resource is believed to last. We can take the weight of a feature *j* (see feature engineering section above), and we can try to understand how a unit increase in such as feature produces a change of $\hat{\beta}_j$ units to the output *Y*. However, the problem with this model is that it considers features to influence the outcome linearly and does not consider interactions between features unless we explicitly add them to the model.

Random Forest aims at solving some of the shortcomings of linear models. In particular, they are based on a combination of classifiers or regressions known as decision trees, and therefore produce non-linear relationships between the features and outcome. However, because decision trees tend to overfit the data, Random Forest bootstraps the data to reduce the model's variance. A further improvement that Random Forest applies is to randomly sample the features themselves, essentially making each of the trees substantially different from each other compared to simple bootstrapping. The effect of such *bagging* (i.e., bostrapped aggregation) and feature bootstrapping is that the variance decreases with more trees. Random Forests have been shown to work remarkably well without much hyperparameter tuning (James et al., 2013).

*Model interpretability*

Even though the Tobit model is highly interpretable, it is challenging to compare the weights of that model to the feature importance from the Random Forest. We use SHAP values to compare the feature importances in our models. SHAP importances can fill this gap. Roughly speaking, SHAP values estimate how much a feature makes a classifier or regression change its prediction compared, on average, to all possible combinations of their features (Lundberg et al., 2020). Importantly, we can apply this definition to any model. Therefore, we can provide some clues as to which features are more critical in our predictions, whether using the Tobit model or Random Forest.

# Results

## *Prediction performance*

In our dataset, we have 58,871 resources shared as URLs within scientific publications. The average lifespan of these resources is 19.45 years (Fig. 1). Because we have the entire survival history of each URL, we can attempt to reproduce the predictability of survival presented in (Zeng et al., 2019). Indeed, by using the features presented in our feature engineering section (see Methods), we estimate a result similar to the conclusion in that article: the URL domain is one of the factors that determine the longevity of a web resource and the predictability of a resource being unavailable in (Zeng et al., 2019).

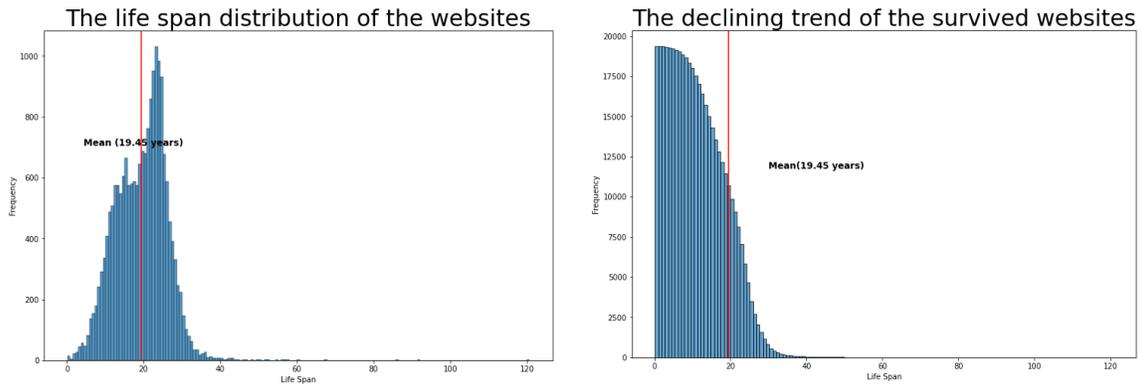

**Figure 1** Life span distribution of our dataset. Left panel: distribution of life span. Right panel: the probability of a resource having more than a given life span. The mean life span is 19.45 years.

We then wanted to understand the simple correlations between the features proposed in Methods for both resources that are available today and those that are not. We found that for both the unavailable and available resources (Fig. 2 and Fig. 3, respectively), the features corresponding to citations are positively correlated as expected. However, the features are less correlated in the available resources (Fig. 3) correlations, suggesting that they are more diverse.

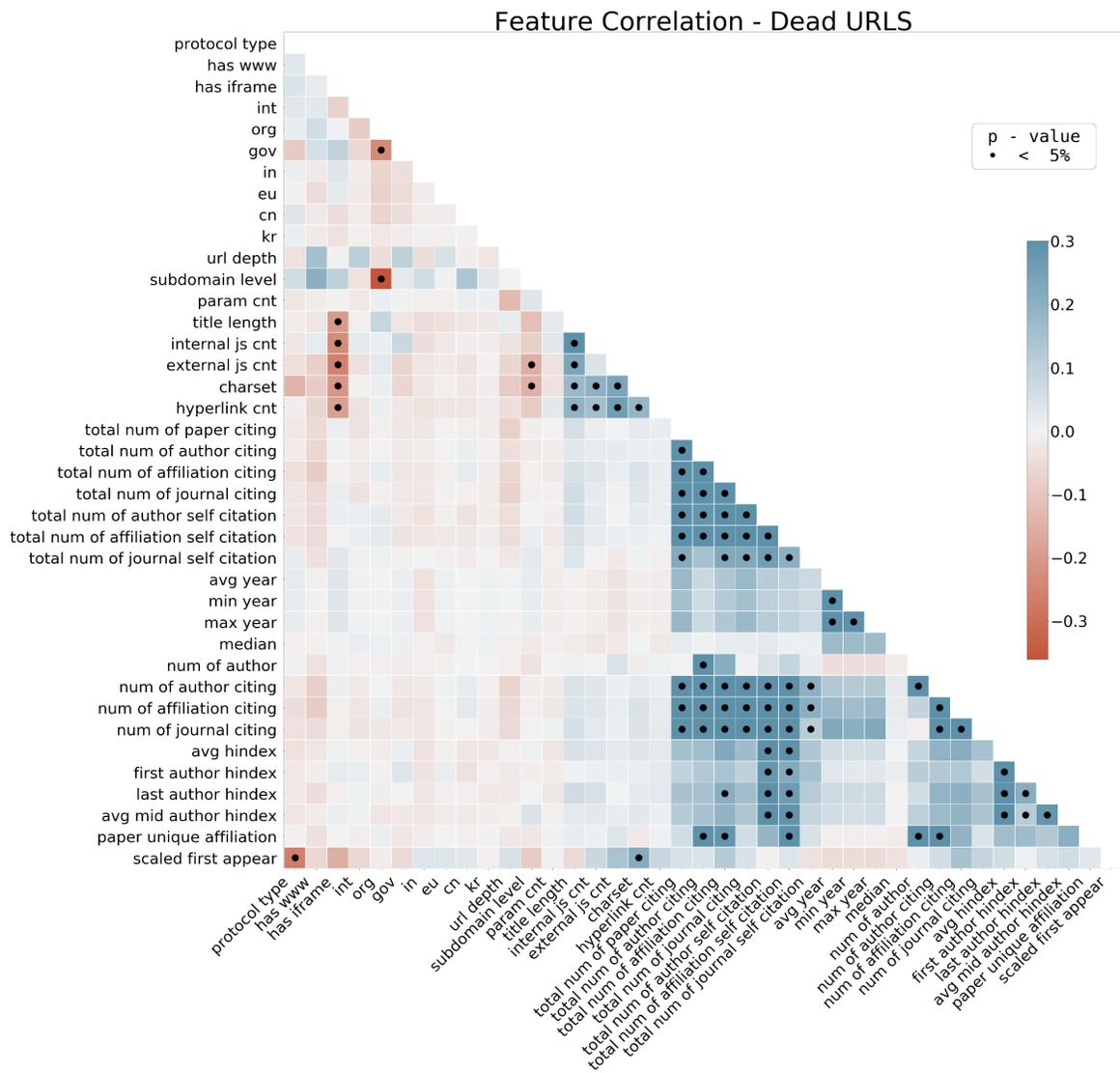

**Figure 2. Correlations between features and their significance values for unavailable resources.** Naturally, there is a heavy positive and significant correlation among features related to citations and references.

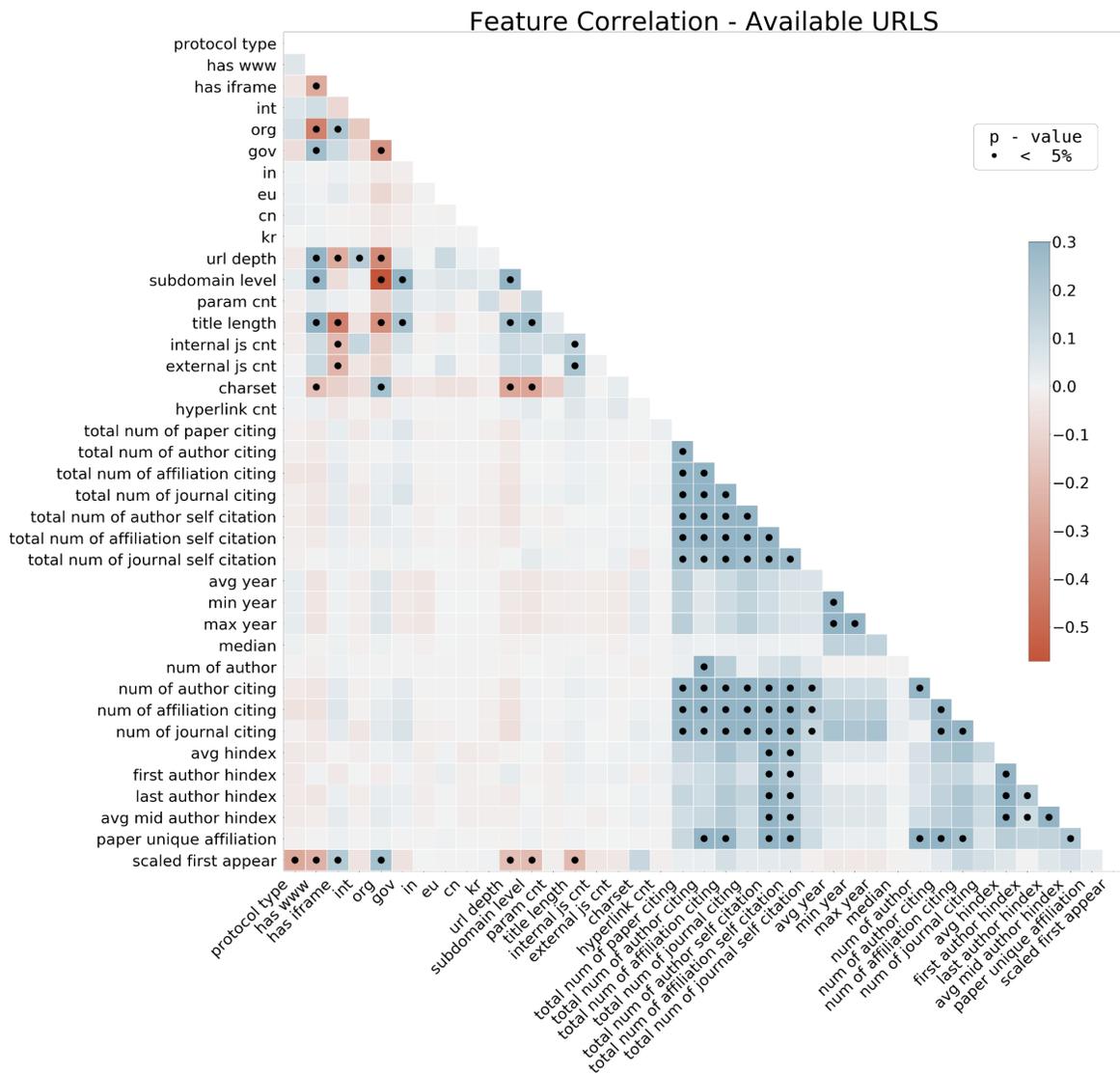

**Figure 3. Correlations between features and their significance values for available resources.** Similar patterns compared to unavailable resources (Fig. 2), but with lower correlation values.

We then wanted to understand how we can predict the longevity of resources for resources that disappeared in the past. For doing so, we simply tried a linear regression with elastic net regularization, which resulted in an estimated validation R squared of 0.164. However, our data is censored (i.e., we do not have information about resources that are still alive), so we need the more sophisticated Tobit model presented in the methods section.

In our dataset, a high percentage of the resources are still available today (90.49%). Still, we need to account for those for which we do not know when they disappear. In this case, we cannot compute the performance using the traditional formulation of $R$ squared, but instead, we have to do it with pseudo-R-square

$$pseudo\ R^2 = 1 - \frac{\log p(Y|\beta)}{\log p(Y)},$$

where $p(Y|\beta)$ comes from the Tobit model. With this model, we estimate our pseudo-R-squared to be 0.045 in cross-validation. The pseudo R squared and the R squared cannot be compared because the units are much different. Finally, pseudo R squared parameters based on loglikelihood tend to be much lower than traditional R squared because loglikelihoods can be much bigger numbers compared to traditional variance (Mbachu et al., 2012).

***How a resource is shared but not who shares it is important for longevity.***

We now analyze what the Tobit model estimated to be important in the prediction. To estimate the uncertainty about each of the models, we bootstrapped the parameters as well (see Methods)[2]. Because we standardized the features, we can, in principle, compare them one to another (Fig. 2). Using this idea, we conclude that features related to technology (charset, parameter count, and domain for positive factors; a locally-hosted javascript library represented by feature `internal_js_cnt` and whether it has an iFrame) are contributing factors to the prediction. On the other hand, we did not find strong predictability of other factors such as the $h$ index, number of citations and references, and the journal's prestige. These results suggest that technology plays a significant role in making sure that resources stay alive longer than average.

---

[2] Our model also has a regularization parameter and we can see how the parameter changes as the amount of regularization increases, a plot known as "regularization path" (Fig. S1, Supplementary Material)

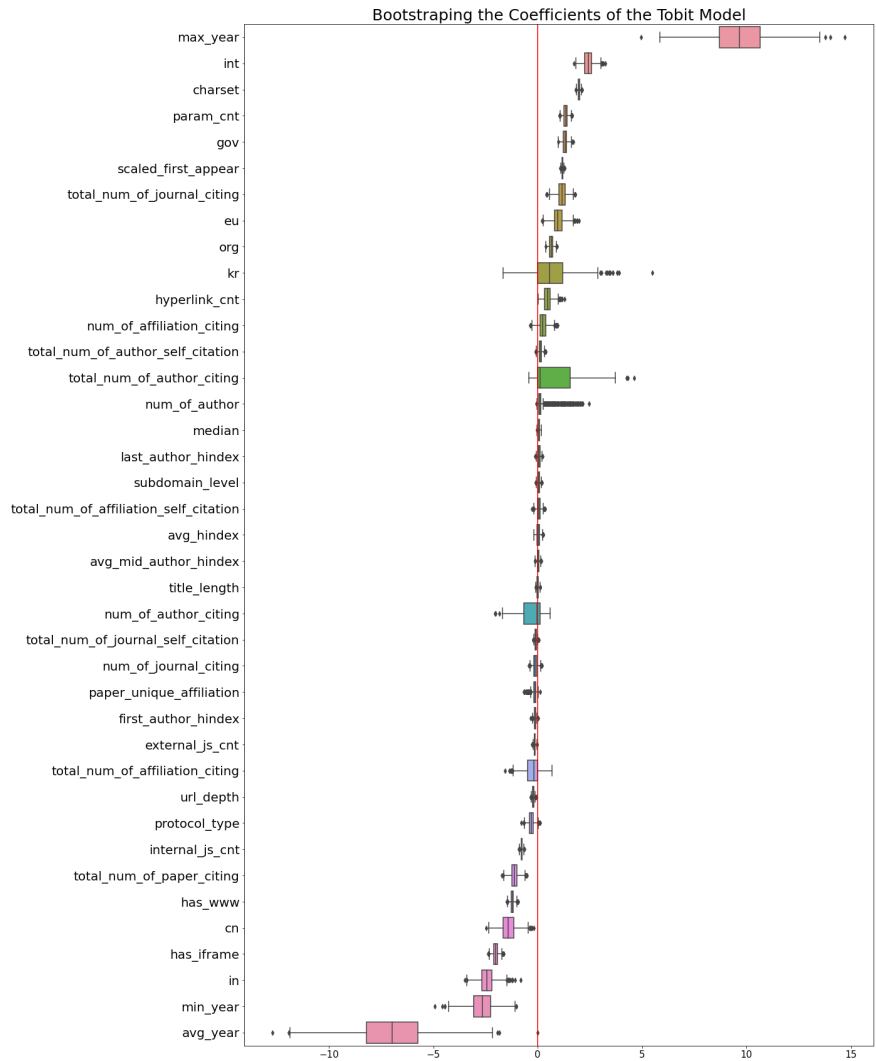

**Figure 4.** Feature importance for predicting the longevity of resources shared by scientific articles.

One of the issues with the Tobit model we proposed is that it does not capture potential non-linear relationships between the features and the longevity of the data. To account for this, we also use a Random Forest model and compare the features that it found relevant to those found significantly different from zero in the Tobit model. Indeed, the coefficients that the Random Forest thinks are most important are highly related to the Tobit model (Fig. 5), including the use of an old Javascript library (`internal_js_cnt` feature) and whether it uses modern HTML structure (`charset` feature).

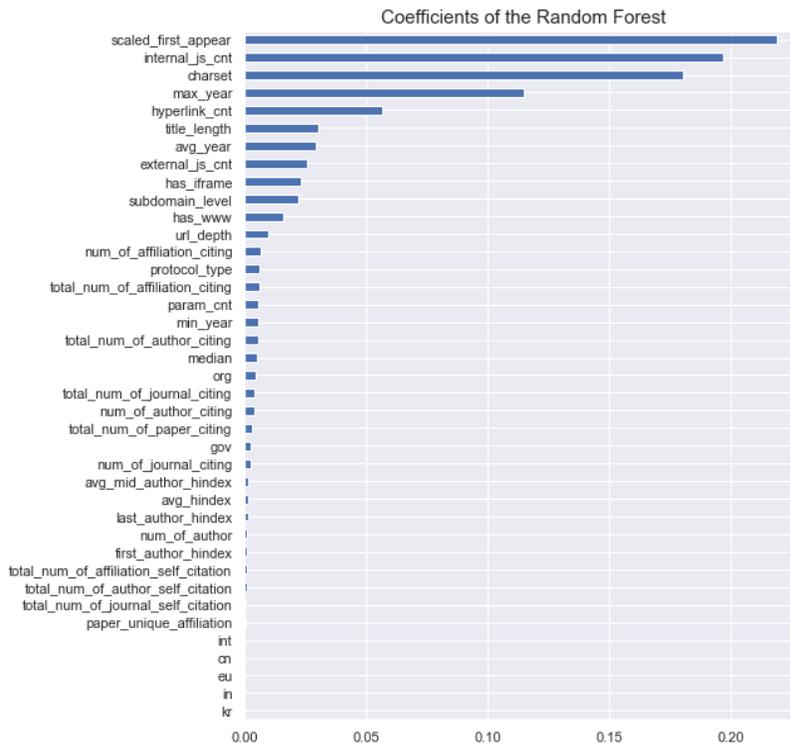

**Figure 5.** Feature importance from Random Forest. The essential features are mostly related to the technology used for sharing data.

One of the issues with feature importance in Random Forest is that it does not inform us in which direction the prediction is happening. We use the SHAP values to do this additional analysis (see Methods, Fig. 6, right panel). This analysis, in principle, allows us to compare both models and see if they make similar predictions. Indeed, we find that the SHAP value of the Tobit model is similar too (Fig. 7, left panel). In both of them, technical-related features are still at the top of the feature list. Unlike the Tobit model, the SHAP values of many features in Random Forest are mixed with the lower value. Only the charset and `internal_js_cnt` have a polarity effect over the average prediction. The feature scaled_first_appear is the top contributor pushing up the average prediction, followed by three features max_year, internal_js_cnt, and title_length. Like the SHAP value in the Tobit model, the technical-related features contribute most to the average prediction.

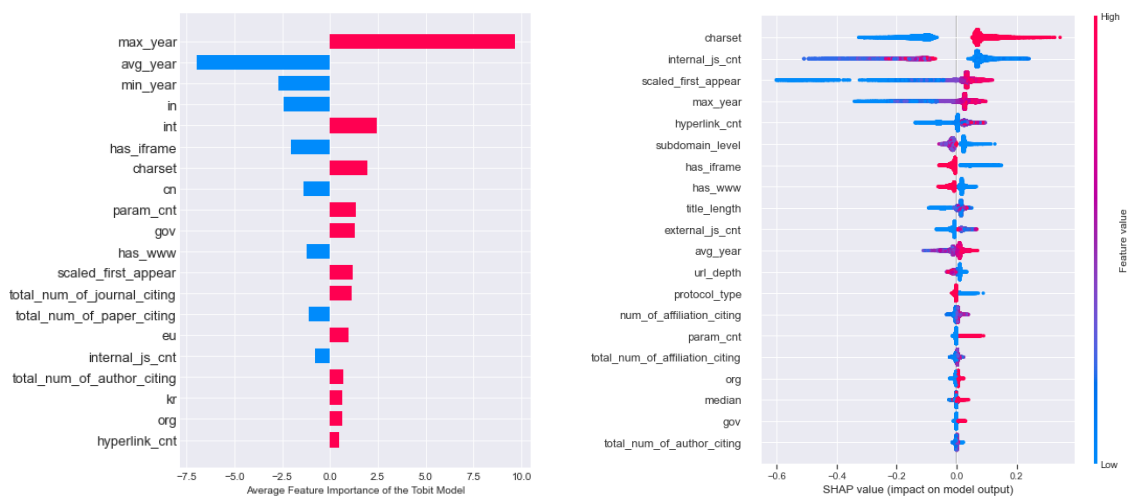

**Figure 6.** SHAP feature importance for Tobit model (left panel) and Random Forest (right panel)

## Discussion

This research aims to understand the factors that predict the longevity of resources shared in scientific publications. First, we use a large sample of open access publications to obtain the URLs of such resources. Then, we reconstructed how these resources looked like throughout time by using archive.org's Wayback Machine. This recovery allowed us to understand the factors of the technology used when the authors shared the resource (e.g., type of HTML or Javascript used). Finally, we crosslinked the publication information with the author's references, affiliation, and journal's prestige. Overall, we found that the factors relevant in the prediction were mostly related to the technology used to share the resources more than prestige. We now discuss our findings of our research, limitations, and future work.

The feature importance analysis provides interesting clues about factors related to longevity (Figures 4, 5, and 6). First, the lifespan of a resource is predicted to be shorter if the article that shares it has many self-citations. While self-citations are not a negative factor per se (Kulkarni et al., 2011), it is a reasonable finding as unusually self-cited work might not be used as much by others. Therefore, there is no incentive to keep it current and available. Second, the count of hyperlinks in the HTML and the depth of the URL behave quite identically. A longer URL will likely have a more complex structure, and we would expect it to be harder to maintain. Also, we found that the iframe tag is negatively associated with longevity. This trend is reasonable because the iframe tag is considered an outdated design that is hard to maintain and prone to malicious payloads (Provos et al., 2008). Finally, the categorical feature charset is negatively correlated to the longevity of a web resource.

Even though we said that factors related to prestige were not comparatively as crucial as the technology, it is still helpful to try to understand them. We found, for example, that the number of different affiliations cited is more critical than the number of other scientists being cited or the prestige (e.g., h-index) of the authors. This trend suggests that the more we cite institutions instead of individual scientists, the more likely we are to care about the resources we are citing—publications with more institutions are cited more often (Gazni & Didegah, 2011), although some have found inter-collaboration is also important (Fu et al., 2018). This fact may influence how much we care about the resources, influencing their maintenance and relevance. In the future, we will explore in more detail these factors through detailed analysis of citations to affiliations vs. individuals.

Previous research has found some similar trends about resource availability. In the work of (Zeng et al., 2019), the authors found that certain domains such as the India, and China are more likely to have resources unavailable and that domains such as .org and .gov are more related to resources being available. Contrary to Zeng & Acuna, 2019, we found that European Union is positively associated with longevity. Still, we think the first set of factors is related to technology use trends in Indian and Chinese higher education institutions, which might be lenient at the moment. Evidence in information systems suggest that internet citations are becoming more and more prevalent in both countries (Chen et al., 2014; Milham et al., 2018; Sampath Kumar & Prithvi Raj, 2012). Conversely, the resources shared in .gov domains tend to be governmental institutions, which, comparatively, may have more established standards and resources for data management. Similarly, domains such as .org might have similar better established governance. Future research will explore data management policy and archiving across countries and types of institutions.

There are several limitations to our research. First, we are analyzing publications in biomedical sciences. Perhaps the technology used is different in more technically sophisticated fields such as engineering and computational science. Also, the factors that affect resources being shared might not be the same. Previous research, for example, showed that scientists are very used to sharing data and code through systems such as figshare.com, GitHub and Amazon's AWS S3 cloud storage (Plantin et al., 2018). However, it has also been found that this research might not be reproducible, even though the resources are available. For example, the data might lack usage instruction, the code might not run, or the results might not match what is reported in the original article (Peng, 2011). It is worth noting that the same could happen in our sample of data. For example, even if we find that resources are "available," we do not know if they are usable. Previous research has found that Excel files shared alongside scientific articles, for example, might contain formulas with errors (Ziemann et al., 2016). Similar problems could be present in biomedical sciences. Future research will explore this subtle but harder-to-examine difference.

We are only limiting our analysis to scientific journals rather than conferences or other types of events. However, because conferences tend to present results faster (Valcher, 2015), we would expect that their resources would be shared with less priority, ultimately affecting their ability to be maintained and available. Many initiatives are trying to make resource sharing an essential part of knowledge production (Baker & Millerand, 2010; Kaye et al., 2009; Mannheimer et al., 2019). Significantly, this resource-sharing impetus does not depend on the kind of venue the resource is shared in. In future research, we will explore how the type of venue is vital for such prediction. Unfortunately, many conference publications are not included in the major citation index or database (e.g., available in Web of Science, Scopus, or the Microsoft Academic Graph). Therefore, we expect that the data about these other kinds of events is more scarce.

Our work examined the essential dimension of longevity in resource sharing. To the best of our knowledge, it is one of the first investigations of this issue. Our work shows that technological factors are perhaps more critical than previously thought. In a sense, our results show that the monetary or other advantages by the authors and institutions are not as important as the technology where the resources are shared. Of course, both factors are correlated—high-resource institutions will help store resources with better technologies. Still, our results might inform how institutions, funding agencies, and journals might approach sharing resources in the future. For example, we could add technological requirements for sharing information. In this sense, the guidelines provided by institutions such as the National Institutes of Health (National Institues of Health, 2020; National Institutes of Health, 2003) make this more explicit. The kinds of analysis we present here can support how these decisions are made in other institutions worldwide. Reproducibility and replicability hinge on our ability to share the resources in the best possible manner (Open Science Collaboration, 2015). The factors presented by our study shine light on what could be the steps needed to achieve this ideal.

## Conclusion

The goal of this article was to investigate the factors associated with resource-sharing longevity. We used a large sample of URLs from open access publications and associated them with the web page where they are shared and the factors associated with publications that shared them. We found that the technology used by the sharer is an essential factor. Though significantly less important, the second set of factors was associated with the number of institutions cited by the work sharing the resource instead of the author and journal numbers or prestige. We discussed some limitations, such as only being focused on biomedical publications and journals. Still, we found that our findings could be important for encouraging better technologies for storing resources.

We have many future avenues for exploration. The first is to better understand whether resources available are usable. For example, some resources might be available but contain incorrect data or code that does not run. This difference will require a significantly more sophisticated analysis of the *content* of the resources being shared. Also, we will expand the analysis to conferences, which we expect to suffer from more issues and have shorter longevity.

We hope that our results illuminate a critical component of reproducibility and replicability. Until now, most work has been encouraging resource sharing without regard to how long they are available. Our work helps the scientific community understand the factors that make longevity necessary. Also, it is striking how important technology is for longevity, which might be related to problems of equity around the world regarding access to the latest methods of sharing data. We hope that these issues get increasingly addressed as we achieve the mission of making resources outside manuscript first-class citizens in the production of knowledge.

## Statements and declarations

The authors declare no competing interests.

# Acknowledgments

DEA and JJ were supported by US's Office of Research Integrity grant ORIIIR190049. TZ, LL, and HZ were supported by the National Science Foundation grant #1933803. TZ was also supported by the China Scholarship Council Award #201706190067.

# Supplementary materials

**Table S1. Features and their rationale.** Some of the features are borrowed from (Zeng and Acuna, 2019)

| Feature | Description | Purpose of the feature |
|---|---|---|
| 1. Protocol type | It's the type of the protocol stated in the URL, it can be Http or https. For other protocols like FTP are not included in the analysis. | Since the lifespan of the SSL resource needs a renewal every other 27 months, a web resource with HTTPS protocol will need more labor resources to maintain its availability. An organization may take it into consideration in the first place. On the contrary, the research result will disappear along with the expiration of SSL due to the lack of project plan and maintenance. |
| 2. Depth of the path in a URI | The URI depth is related to the design of the web page. For example, a web page with a REST architecture style uses the path hierarchy to identify the location of a resource. | The depth of the URI represents a hierarchical structure in a web domain. Eg, the REST API is a well designed URL pattern that publishes the resources of an application to the outside in a graceful manner. An easy-to-remember URI is beneficial to the lifespan of a web resource. |
| 3. www or non-www URI | Www is one of the subdomains of a website. From the user's perspective, it is the synonym of the front page which content is bonded to the primary domain. The website developer usually routes the www subdomain to its primary domain by sending a redirect HTTP request. The web resource may not be available without proper maintenance. | From the perspective of SEO, a domain with or without www will have no difference being included in the search engine. However, to unify the content of these 2 domains requires extra effort, because the www suffix is the subdomain of the main domain. The website engineer needs to redirect (HTTP code 300 family) from one to another. This feature will help us to figure out the relationship between the maintenance of the website and its lifespan. |
| 4. The level of the subdomain | Self-explained feature | |
| 5. The number of query parameter in the URI | How many parameters used in the HTTP Get request. | The HTTP Get request accepts additional query parameters in the URL. The argument list is tied to the design of the backend system. The lack of the sustainability of the system design is detrimental to the availability of a web resource. Because referred links in a paper cannot be revised after published |
| 6. Domain suffix with org | A feature from previous research. | From (Tong & Acuna , 2019), the org and int suffix are said to be one of the important factors that positively correlated to the lifespan of the website. |
| 7. Domain suffix with int | A feature from previous research. | |
| 8. Domain suffix with jp | A feature from previous research. | According to the research, the top non-US tools from Japan maintained a high reputation and were frequently cited. It is worthy to explore the suffix with JP to see how its impact on the lifespan of the web resources. |
| 9. Domain suffix with gov | A feature from previous research. | As above |
| 10. Indian domain (.in) | A feature from previous research. | As stated in (Tong & Acuna , 2019), Resources with in/cn/eu/kr negatively contribute to the longevity of a website. |

| # | Feature | Description |
|---|---|---|
| 11. Domain suffix with cn | A feature from previous research. | |
| 12. Domain suffix with eu | A feature from previous research. | |
| 13. Domain suffix with kr | A feature from previous research. | |
| 14. Is the web resource explicitly accessed through port | A flag to identify the existence of the port information in the web resource. | By default, a direct access to a website implicitly goes into the port 80. A URI with explicitly port access may not be a user-friendly design and could be revised in a later version. |
| 15. The size of the HTML source code (in string length) | Self-explained feature | The size of the HTML code is one of the determinant |
| 16. The length of the HTML title in the source code | Self-explained feature | The HTML title abstracts the details of the web resource, which increases the readability of the website. |
| 17. The number of internal JavaScript file | Self-explained feature | While hosting the JavaScript file inside a domain would increase the complexity of the maintenance process, it may give the full autonomy on extension development of the website. We include this feature because how this property affects the lifespan of the web resources is still unclear to us. |
| 18. The number of external JavaScript file | A JavaScript library imported from the location hosted by the same web domain | A JavaScript library imported from the location outside the web domain. Whether this exterior JavaScript resource is accessible or not, its availability relies on the external resource itself. |
| 19. The charset specified in the HTML source code | A categorical variable that defines the type of the char encoding specified in the web page. | A charset classifier is controlling the encoding of the page. The content of the web page is vulnerable to be obfuscated if the decoding charset in the browser is inconsistent with the one specified in the web page. Since the UTF-8 encoding is widely used across many kinds of websites, we believe using UTF-8 encoding will help to extend the lifespan of the web resource. |
| 20. HTML or HTML5 | An indicator to distinguish the HTML5 page from the regular HTML. | The first html5 opened to the public was created on 22 January 2008. So the HTML5 indicator can be considered as a substitution of chronology information.<br><br>While the new features of HTML5 equipped the webpage with the ability of presenting a variety kinds of information, processing the unstructured data in the webpage requires more effort to be utilized by the third party. |
| 21. The iFrame tag usage indicator | Whether the iFrame tag is used in the web page or not | The iFrame tag can be used to control the layout of the page, or to address the Cross-origin resource sharing issue.<br>The single-page application design is extensively used in the modern website. If the source code of a website utilized the iframe, then this page could be built by the technology of the previous generation. So this feature is chronological related |
| 22. The number of hyperlink on the web page | The number of web links that represent other resources on the internet. | Publishing the information through the web link is a complementary way of sharing the knowledge. Extensively using web resource or traditional text conducted article sharing are different ways of researching. Including this feature helps to identify how the way of page organizing impacts the longevity of the web resource. |
| 23. Total number of publications referenced by | Self-explained feature<br><br>total_num_of_paper_referencing | The paper consolidates the information of other cited papers. The referred material is the extension of the paper. This feature helps to measure how the paper referral impacts its lifespan. |
| the paper | | |
| 24. Total number of authors referenced by the paper | Self-explained feature<br><br>num_of_author_referncing | Researchers have their own study field. Although we cannot claim the paper is an interdisciplinary research by solely relying on the number of the author, the number does tell us the degree of the innovation of the paper. |
| 25. Total number of affiliations referenced by the paper | Self-explained feature<br><br>num_of_affiliation_referencing | The study area among researchers in the same affiliation is closer than that in the different affiliation. Collaborating this feature with the total number of referred affiliation in the analysis helps to explore how the concentration of the research field contributes to the longevity of resources. |
| 26. Total number of journals referenced by the paper | It measures the diversity(number) of the different referred journals.<br><br>num_of_journal_referencing | |
| 27. The total number of self-citations in a paper. | The number of cited paper belongs to the authors themselves<br><br>total_num_of_author_self_citation | This feature evaluates how strict the author of this paper is on the previous research direction. |
| 28. The number of citations of a paper that is published by the same affiliations. | Self-explained feature.<br><br>total_num_of_affiliation_self_citation | The communication among researching is more feasible to carry out in the same affiliation. But it is vulnerable to get a homogeneous result if the paper refers to more resources from the same affiliation. |
| 29. The number of citations of a paper that is published on the same journals. | Self-explained feature.<br><br>total_num_of_journal_self_citation | Each journal has its unique criteria of evaluating and accepting the paper. It reflects the research quality of the paper. Gauging this number helps to measure the impacts of diversity of the quality to the lifespan of web resources. |
| 30. The average year of publications referenced by the paper | Self-explained feature.<br><br>avg_year | A chronology information that represents the overall years of the cited paper. |
| 31. The earliest year of publications referenced by the paper | Self-explained feature<br><br>min_year | As above comment |
| 32. The latest year of publications referenced by the paper | Self-explained feature<br><br>max_year | As above comment |
| 33. The median year of | Self-explained feature | As above comment |
| publications cited by the paper | median | |
| 34. The number of authors | How many authors collaborated on this paper.<br><br>num_of_author | Behind this number is the effort spent by each author. It indirectly reflects the scale of the paper in terms of the authors. |
| 35. The number of affiliations that published the paper's citations. | How many affiliations are engaged in the citations.<br><br>paper_unique_affiliation | |
| 36. The total number of authors of citations of a paper. | How many authors(with duplication) wrote the citations of a paper.<br><br>total_num_of_author_citing, | Behind this number is the effort spent by each author. It indirectly reflects the scale of the paper in terms of the resources it used. |
| 37. The total number of journals that published the paper's citations. | total_num_of_journal_citing | |
| 38. The total number of affiliations that published the citations of the paper | How may affiliations (with duplication) are involved in this particular paper<br><br>total_num_of_affiliation_referencing | This feature measures the degree of the collaboration among affiliations. |
| 39. The arithmetic average level of H-Index | It is the number averaged by all years of h-index among all the authors of a paper.<br><br>avg_hindex | These features use the H-index of the authors to estimate the longevity of resources. |
| 40. The level of H-Index of the primary author | Self-explained feature<br><br>first_author_hindex | |
| 41. The level of H-Index of the last author | Self-explained feature<br><br>last_author_hindex | |
| 42. The level of H-Index of authors other than the first and last. | Self-explained feature<br><br>avg_mid_author_hindex | This feature evaluates the academia degree of the minority contributors of this paper. |
| 43. The end year of the website | The last available day of the web resource | We scaled the dependent variable by using the number 1990 to subtract the year it was last available, because the first web page of the world was |

| | | published in 1991. |
|---|---|---|

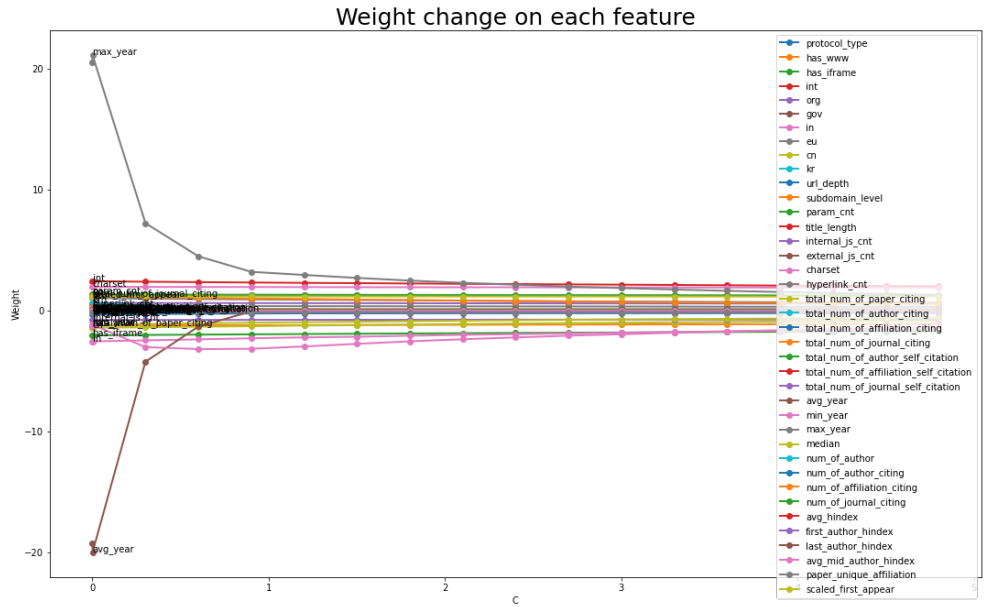

**Fig. S1. Regularization path of Tobit regression**